\documentclass[11pt]{article}
\usepackage{authblk}
\usepackage{graphicx}
\usepackage{amsmath}
\usepackage{amssymb}
\usepackage{hyperref}
\usepackage{csquotes}
\usepackage{geometry}
\usepackage{titlesec}
\usepackage{etoolbox}
\usepackage{placeins}
\usepackage{upgreek}

\usepackage[
  backend=biber,
  style=numeric-comp,
  sorting=none,
  doi=true,
  url=false,
  isbn=false,
  eprint=false
]{biblatex}

\titleformat{\section}
  {\normalfont\Large\bfseries}{\thesection}{1em}{}
\titleformat{\subsection}
  {\normalfont\large\bfseries}{\thesubsection}{1em}{}
\titleformat{\subsubsection}
  {\normalfont\normalsize\bfseries\itshape}{\thesubsubsection}{1em}{}

\title{Persistent Homology-Based Indicator of Orientational Ordering in Two-Dimensional Quasi-Particle Systems Applied to Skyrmion Lattices}
\author[1, 2]{Michiki Taniwaki}
\author[2]{Thomas Brian Winkler}
\author[2]{Jan Rothörl}
\author[2]{Raphael Gruber}
\author[3]{Chiharu Mitsumata}
\author[1]{Masato Kotsugi}
\author[2]{Mathias Kläui\thanks{Corresponding author: Mathias Kläui}}
\affil[1]{Department of Materials Science and Technology, Tokyo University of Science, Niijuku 125-8585, Japan}
\affil[2]{Institute of Physics, Johannes Gutenberg University Mainz, 55099 Mainz, Germany}
\affil[3]{Graduate School of Pure and Applied Sciences, University of Tsukuba, Tenodai 305-8571, Japan}

\date{\today}

\begin{document}

\maketitle

\vspace{-1em}
\begin{abstract}
    \noindent
    Two-dimensional (2D) particle systems, such as magnetic skyrmions, exhibit topological phase transitions 
    between unique 2D phases.
    However, a simple and computationally efficient methodology to capture lattice configurational properties 
    and construct an appropriate, easily calculable descriptor for phase identification remains elusive.
    Here, we propose an indicator for topological phase transitions using persistent homology (PH). 
    PH offers novel insights beyond conventional indicators by capturing 
    topological features derived from the configurational properties of the lattice. 
    The proposed persistent-homology-based indicator, which
    selectively counts stable features in a persistence diagram,
    effectively traces the lattice's ordering changes, as confirmed by comparisons with the conventionally used 
    measure of the ordering (the magnitude of the orientational order parameter $\langle|\Psi_6|\rangle$), 
    typically used to identify lattice phases. 
    We demonstrate the applicability of our indicator to experimental data, showing that 
    it yields results consistent with those of simulations. 
    This experimental validation highlights the robustness of the proposed method for real physical systems beyond 
    idealized simulated systems.
    While our method is demonstrated in the context of skyrmion lattice systems, the approach is general 
    and can be extended to other two-dimensional systems composed of interacting particles. 
    As a key advantage, our indicator offers lower computational complexity than the conventionally used measures.
\end{abstract}

\section*{Introduction}
\indent
Magnetic skyrmions are chiral, vortex-like spin textures characterized by topologically enhanced 
stability \cite{ref1,ref2}, often treated as quasi-particles due to their particle-like behavior \cite{ref3,ref4}. 
These structures have garnered significant attention from researchers for their energy-efficient manipulation 
via spin-transfer and spin--orbit torques \cite{ref5}, offering promising pathways toward next-generation information 
storage and unconventional computing devices \cite{ref35}. 
A possible application of skyrmions is the racetrack memory as a high-density memory technology \cite{ref6,ref7,ref36}. 
Such future skyrmionic applications are based on the behavior of multiple skyrmions.
It is, therefore, important to investigate their physical properties and their collective arrangements to achieve 
such technological applications using skyrmions.

From a fundamental physics perspective, ensembles of skyrmions, known
as skyrmion lattices, can exhibit complex phase behavior, including
two-dimensional melting
transitions \cite{ref8}. These transitions
are theoretically captured by the
Kosterlitz--Thouless--Halperin--Nelson--Young (KTHNY) framework, which describes the emergence of intermediate hexatic phases and the role of
topological defects such as dislocations and disclinations \cite{ref8}. 
Although this behavior has been extensively studied in colloidal systems, skyrmion lattices offer 
distinct advantages due to their tunability and controllable dynamics that can be fully captured 
in real time and real space, making them an ideal platform for investigating two-dimensional phase 
transitions \cite{ref10}.

Despite intensive research, a computationally easy methodology for
describing the configurational properties of lattices remains elusive.
Conventional methods (e.g., using the local orientational order parameter $\psi_{6}(r)$ \cite{ref5,ref10,ref11,ref12}) 
are limited by their reliance on ensemble-averaged quantities and are computationally expensive. 
That is because they normally average the quantities encoded to each particle over all the particles to
identify the state of the systems. 

To address this, we introduce a framework based on Topological Data
Analysis (TDA), a concept from algebraic topology used to analyze 
the geometric structure of objects, specifically applying Persistent Homology (PH) to 
extract configurational properties of lattice configurations \cite{ref13,ref14,ref15,ref16,ref17,ref18}. 
The PH provides a multi-scale view of topological
features by analyzing how connected components and loops emerge and disappear across a filtration. 
This approach has seen successful applications in biological and materials systems 
\cite{ref19, ref20, ref21, ref22,ref23}. 
PH can capture and describe the microscopic processes involved
in the phase transitions \cite{ref15}. 
In this work, we apply PH in a particle-based approach, which models skyrmions as interacting quasi-particles and 
focuses on their positional configuration that does not require knowledge of the full spin texture.
This abstraction enables an efficient computational treatment and highlights the essential geometric 
features responsible for phase behavior \cite{ref3, ref4, ref24}.
We propose a topological
indicator \cite{ref25, ref26}, the
Persistent Generator Count with Relative Stability (PGCRS), which
selectively counts only the robust topological features of the persistent diagram (PD) in each homology
dimension, generated by PH. 
PGCRS reliably detect phase transitions in the lattice, 
correlates with conventional measures, 
and offers a significantly reduced computational complexity.
A distinctive aspect of our approach is inverse analysis \cite{ref27}, 
to trace persistent generators back to specific real-space configurations, enabling
a direct physical interpretation of the microscopic structures responsible for topological phase behavior.
This conceptual shift from averaging local order to counting persistent topological features offers a new 
perspective for understanding 2D phase transitions. 
We apply our framework to experimentally acquired skyrmion configurations, 
demonstrating the practical applicability of our method for real physical systems.
While the focus of this work is on skyrmion systems, the underlying method can be extended to other 
two-dimensional particle systems.


\section*{Results and Discussion}
Fig. 1 presents a schematic overview of the analysis workflow. 
We first (a) obtain the coordinates of skyrmions from the experimental data, and then (b) apply the PH analysis 
with (c) the calculation of the conventionally used measure of the ordering. 
Finally, we (d) compare the two indicators.
The experimental procedure is detailed in the Methods section.  
The PH analysis is described in the following subsections and illustrated in Figs.~2--4,  
where the PDs and their interpretation through inverse analysis are provided.  
Subsequently, the PH-based indicators and the conventionally used measure are compared 
in terms of consistency and computational cost, as summarized in Fig.~4 and Table~1.  
Validation using Molecular Dynamics (MD) simulations is presented in the Supplementary Information (Figs.~S1--S6).

\begin{figure}[ht]
    \centering
    \includegraphics[width=0.9\textwidth]{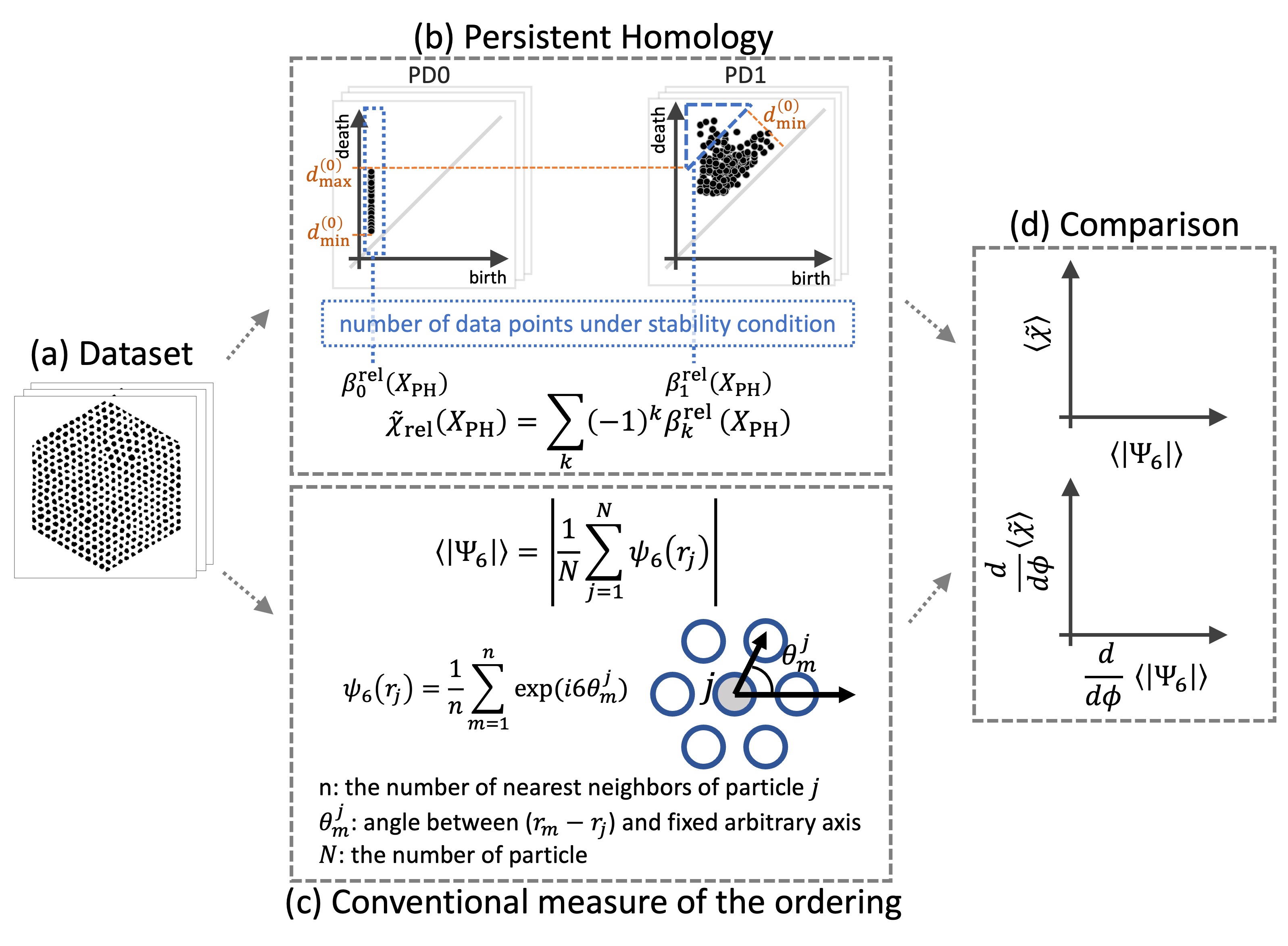}
    \caption{
        Schematic of the analysis workflow:  
        (a) skyrmion coordinates,  
        (b) Persistent Homology (PH) analysis,  
        (c) calculation of a conventional lattice-ordering measure, and  
        (d) comparison between the PH-based indicator and the conventional measure.
    }
    \label{fig:fig1}
\end{figure}

\FloatBarrier
\subsection*{Persistent Homology and Persistence Diagrams for Experimental Skyrmion Lattices}

Fig. 2 illustrates the filtration process and the corresponding persistence diagrams (PDs) for the 0th- and 1st-degree homology dimensions.  
The filtration process is performed by continuously increasing the radius of disks centered at the skyrmion coordinates, 
thereby tracking the emergence and disappearance of topological features such as connected components and loops.  
The birth and death values in the PDs correspond to the specific filtration parameter $r$ 
at which these topological features appear and vanish, respectively.  
The PD$_0$ diagram captures the birth and death of connected components (0th-degree homology), 
while the PD$_1$ diagram records the birth and death of loop-like structures (1st-degree homology).

In particular, the data points in PD0 indicate when individual skyrmions (disks) merge; 
thus, all birth values are zero, and the death values correspond to the connection events.  
The number of data points in PD0 equals the number of skyrmions in the system.  
In PD1, the lifetime ($\text{death} - \text{birth}$) represents 
the persistence of loop-like structures formed by connected skyrmions.  
A large lifetime indicates robust loops that persist over a wide range of the filtration parameter, 
whereas a small lifetime corresponds to transient loops that quickly disappear as the filtration parameter varies, 
or possibly to noise.

For a detailed explanation of the PH filtration process, please refer to Ref.~\cite{ref14}.

\begin{figure}[ht]
    \centering
    \includegraphics[width=0.9\textwidth]{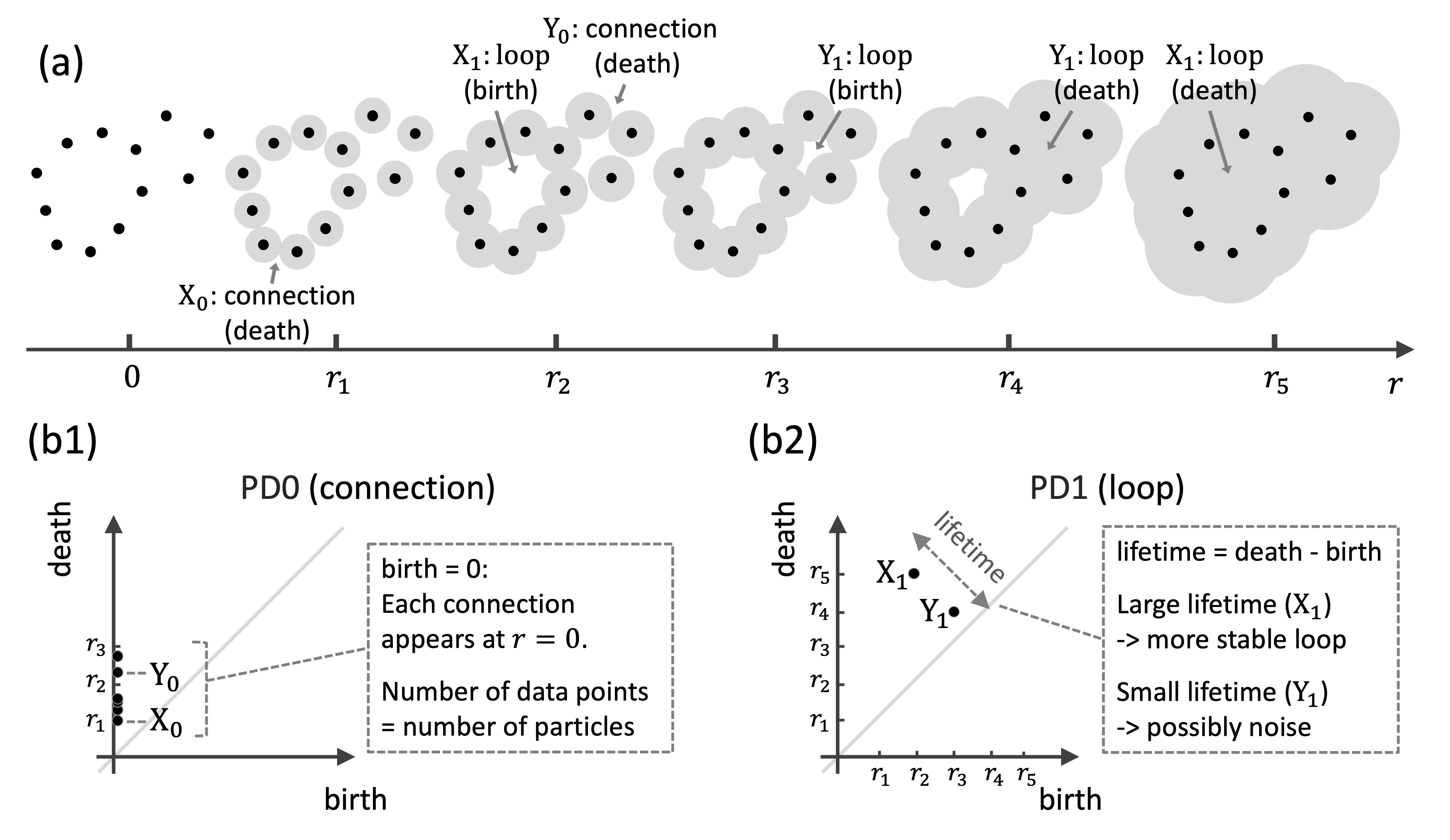}
    \caption{
        Filtration and persistence diagrams (PDs) in the 0th and 1st homology dimensions.  
        (a) Illustration of the topological features---connected components and loops---emerging during the filtration process, 
        where the radius of disks centered at the skyrmion coordinates is varied.  
        (b1) PD0 (0th-degree homology) and (b2) PD1 (1st-degree homology) with interpretations.  
        For a detailed explanation of the PH filtration process, please refer to Ref.~\cite{ref14}.
    }
    \label{fig:fig2}
\end{figure}

\FloatBarrier

Fig. 3 shows the persistence diagrams (PDs) of the 0th- and 1st-degree homology obtained from three states 
of the experimental skyrmion lattice under applied out-of-plane (OOP) magnetic fields of 
$B = 60$, $84$, and $108$~$\upmu$T, corresponding to the solid, hexatic, and liquid phases, respectively \cite{ref10}. 
In each PD, the "Birth" and "Death" values represent the filtration stages during which topological features emerge and disappear, respectively, 
and the color map indicates the multiplicity of generators. 
The "Birth" and "Death" values correspond to the radii of disks grown during the filtration process
, as explained in Fig.~2.
The PD0 for the disordered configuration (e.g., $B = 108$~$\upmu$T) displays a broader 
distribution compared to the more ordered configuration (e.g., $B = 60$~$\upmu$T). 
Additionally, the PD1 for the disordered state exhibits a greater number of generators with large lifetimes (i.e., 
features farther from the diagonal line), while the ordered state tends to show fewer such persistent features.

\begin{figure}[ht]
    \centering
    \includegraphics[width=0.9\textwidth]{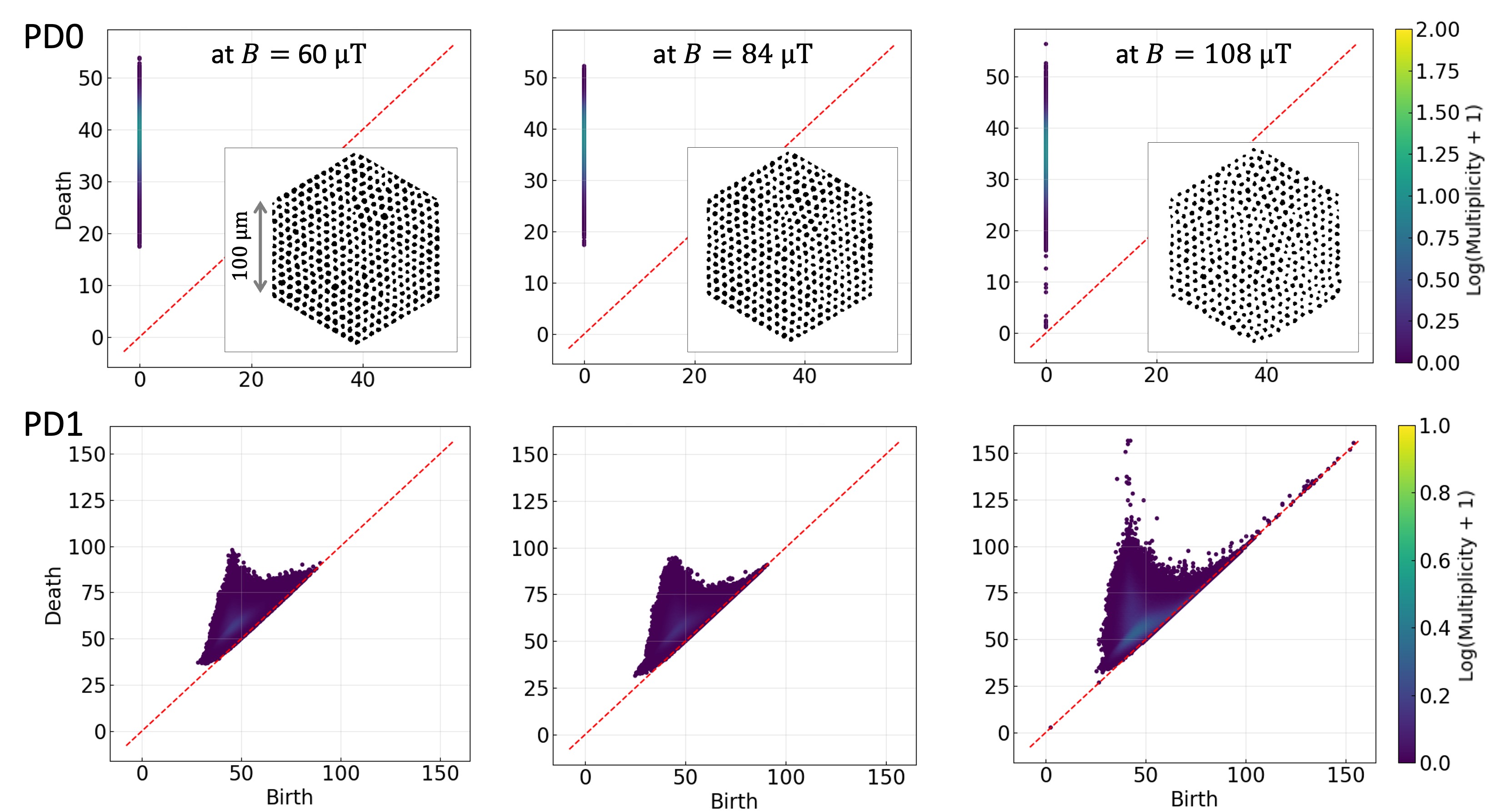}
    \caption{
        Average persistence diagrams (PDs) of the 0th- and 1st- degree homology, for three states under applied OOP magnetic 
        fields of $B = 60$, $84$, and $108$~$\upmu$T, corresponding to solid, hexatic, and liquid phases. 
        The ”Birth” and ”Death” represent the specific times in 
        a filtration process of persistent homology in which topological features
        emerge and disappear, respectively. The color map represents the multiplicity of generators
        (scatter plots) in the PD. 
        Insets in the 0th-degree homology PDs display the corresponding real-space configurations, 
        identified using a machine-learning-based, pixel-wise classification algorithm~\cite{ref33}.
    }
    \label{fig:fig3}
\end{figure}

Fig.~4 presents the inverse analysis \cite{ref27} for two states at applied OOP magnetic fields $B = 60$ and 108~$\upmu$T, 
which conducts the analysis tracing back the specific data points in PD to the original structure in the real-space configuration. 

In each state, the points labeled (a), (b), and (c) correspond to data points in the PD with a large lifetime
(a), small lifetime and small birth value (b), and small lifetime and large birth value (c), respectively. 
It is observed that data points with a large lifetime originate from a complex structure, while those 
with a small lifetime originate from a simple structure, consistent with the simulation data.
The result reflects the distribution of data points in the PDs (Fig.~3), as the
disordered lattices exhibit more complex features compared with ordered lattices.
Hence, the PH framework effectively links the actual configurational structure of the skyrmion 
lattice to the features represented in the corresponding PD, which 
is difficult to capture with the conventional orientational-order measure.

\begin{figure}[ht]
    \centering
    \includegraphics[width=0.8\textwidth]{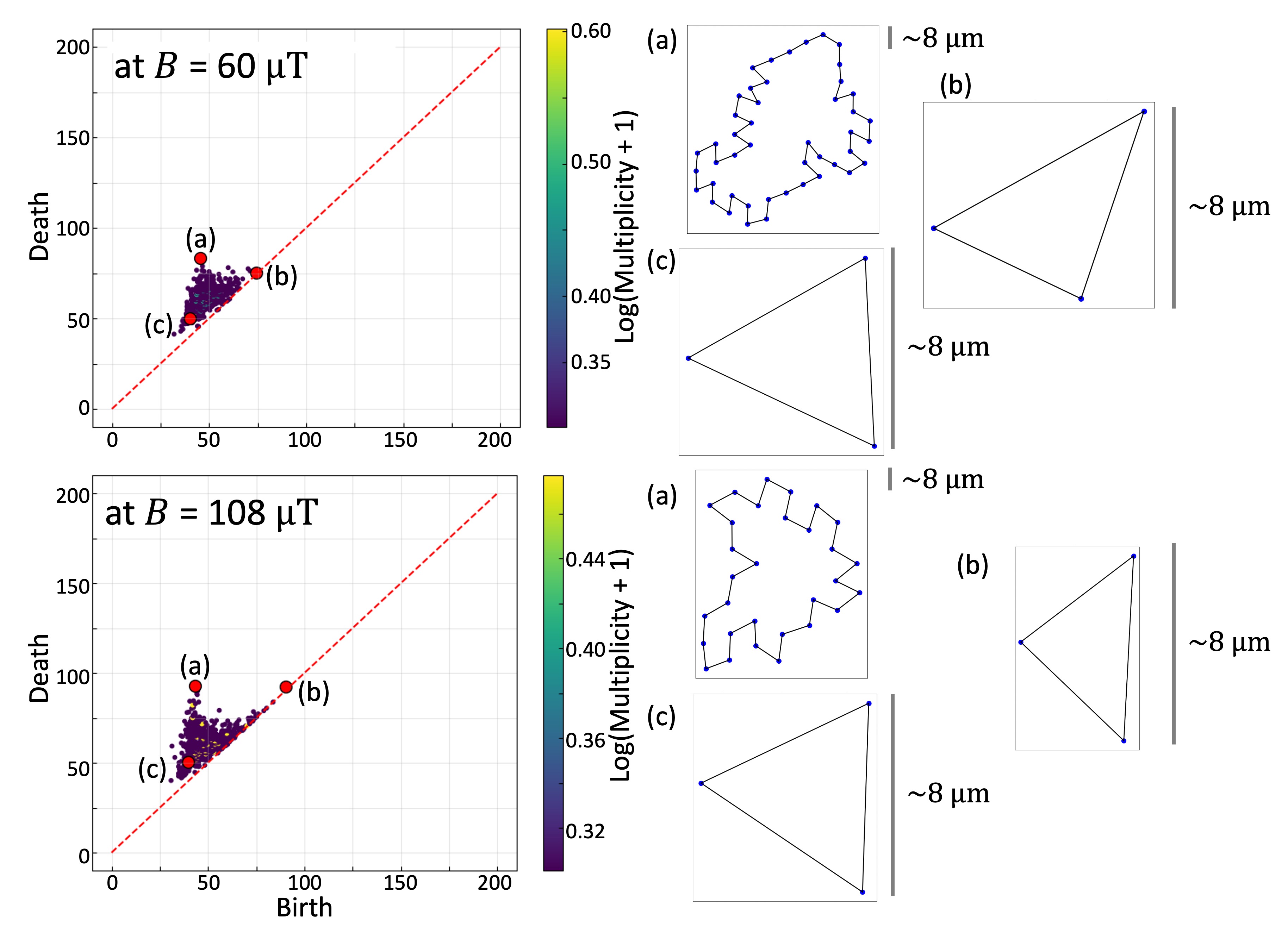}
    \caption{
        Inverse analysis for the two states at $B = 60$ and $108$~$\upmu$T, tracing specific generators
        in the persistence diagram back to real-space configurations. In each state, the points labeled (a), (b), and (c)
        correspond to persistent homology generators with large lifetime (a), small lifetime and small
        birth value (b), and small lifetime and large birth value (c), respectively.
    }
    \label{fig:fig4}
\end{figure}

\subsection*{Persistent Homology-Based Indicator and Comparison with Conventionally Used Measure of Ordering}
Fig.~5 presents the Persistent Generator Count with Relative Stability
as a function of $\langle|\Psi_6|\rangle$, along with the first derivatives of both indicators.
A positive correlation is observed, as confirmed by the Pearson correlation coefficients $r = 0.993$ and $r = 0.861$ for 
the values and their derivatives, respectively. 
The details of the Gaussian Process Regression, used to estimate the first derivatives,
are described in Supplementary information and Fig.~S7.
These results confirm the consistency between the PH-based indicator and the conventional orientational order parameter.
As also discussed in the simulation data, the high sensitivity, in particular the first derivatives, supports the validity of the persistent 
homology-based approach, as phase transitions are characterized by abrupt, non-analytic changes in system properties \cite{ref39}.

\begin{figure}[ht]
    \centering
    \includegraphics[width=0.8\textwidth]{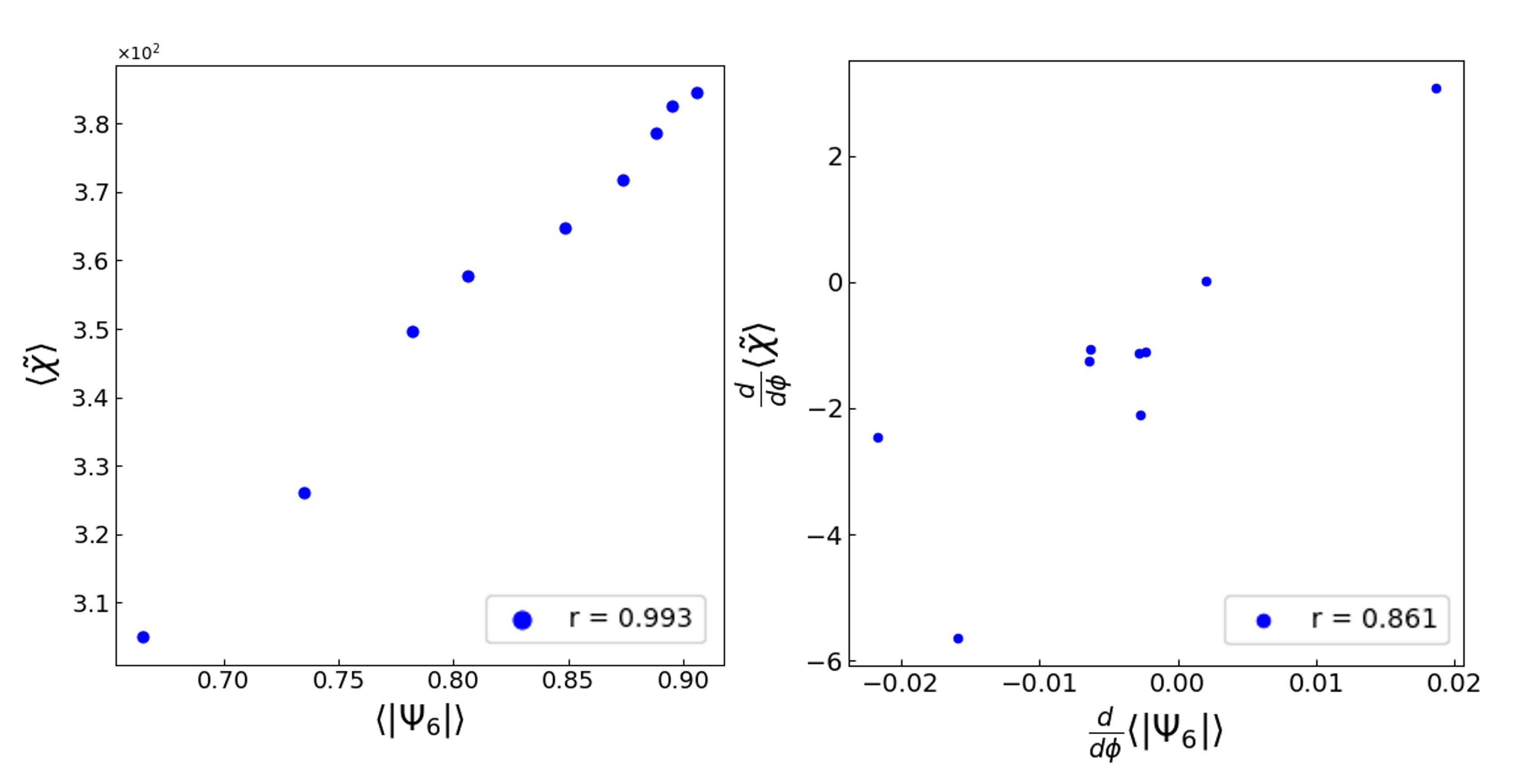}
    \caption{
        Correlation between the Persistent Generator Count with Relative Stability $\langle \tilde{\chi} \rangle$ 
        and the conventional orientational order parameter $\langle|\Psi_6|\rangle$ in the experimental data. First derivatives are 
        compared using Gaussian Process Regression. A good correlation is indicated by the Pearson correlation coefficients $r = 0.993$ and $r = 0.861$.
        }    
    \label{fig:fig5}
    \end{figure}

\FloatBarrier
Finally, we analyze the computational cost of our indicator compared to the conventionally used indicators. 
Table 1 shows the comparison in computational complexity for the
persistent homology-based indicator and the conventionally used measure $\langle|\Psi_6|\rangle$ \cite{ref11, ref31}. 
It is obvious that the newly constructed indicator has achieved
a significant reduction in computational complexity, as the dimension of
the system in this work is 2, and therefore the cost is $O(N)$.
Furthermore, in the simulation data, the actual runtime for the persistent homology-based indicator is 2.34 seconds per frame, 
whereas the conventionally used one requires $4.49 \times 10^{3}$ seconds in our computational environment.

The ability of persistent homology analysis to capture the ordering may stem from the same geometric foundation as 
the orientational order parameter, since both rely on Voronoi tessellation, complemented by algebraic topology techniques \cite{ref38}. 
One limitation, however, is that the persistent homology-based indicator may require comparisons across different system states to 
provide meaningful insights. 
In contrast, the conventionally used measure of the ordering $\langle|\Psi_6|\rangle$ has a clear physical 
interpretation due to its normalization, which ensures values between 0 (disordered state) and 1 (perfect hexagonal order) \cite{ref5}.
Nevertheless, the PH-based indicator can be further generalized through an entropy-based formulation,  
as algebraic topology provides a well-defined correspondence with thermodynamic quantities \cite{ref25,ref26}.  
Such a generalization could offer a unified framework for describing both structural complexity and 
statistical behavior.

\begin{table}[h!]
    \caption{
        Comparison of computational complexity for the
        persistent homology-based indicator and the averaged absolute value of
        the local orientational order parameter ~\cite{ref11,ref31}. Here, $N$
        is the number of particles in the system, and $d$ is the dimension. In
        this work (simulation data), $N = 65000$ and $d = 2$.
    }
    \centering
    \includegraphics[width=0.5\textwidth]{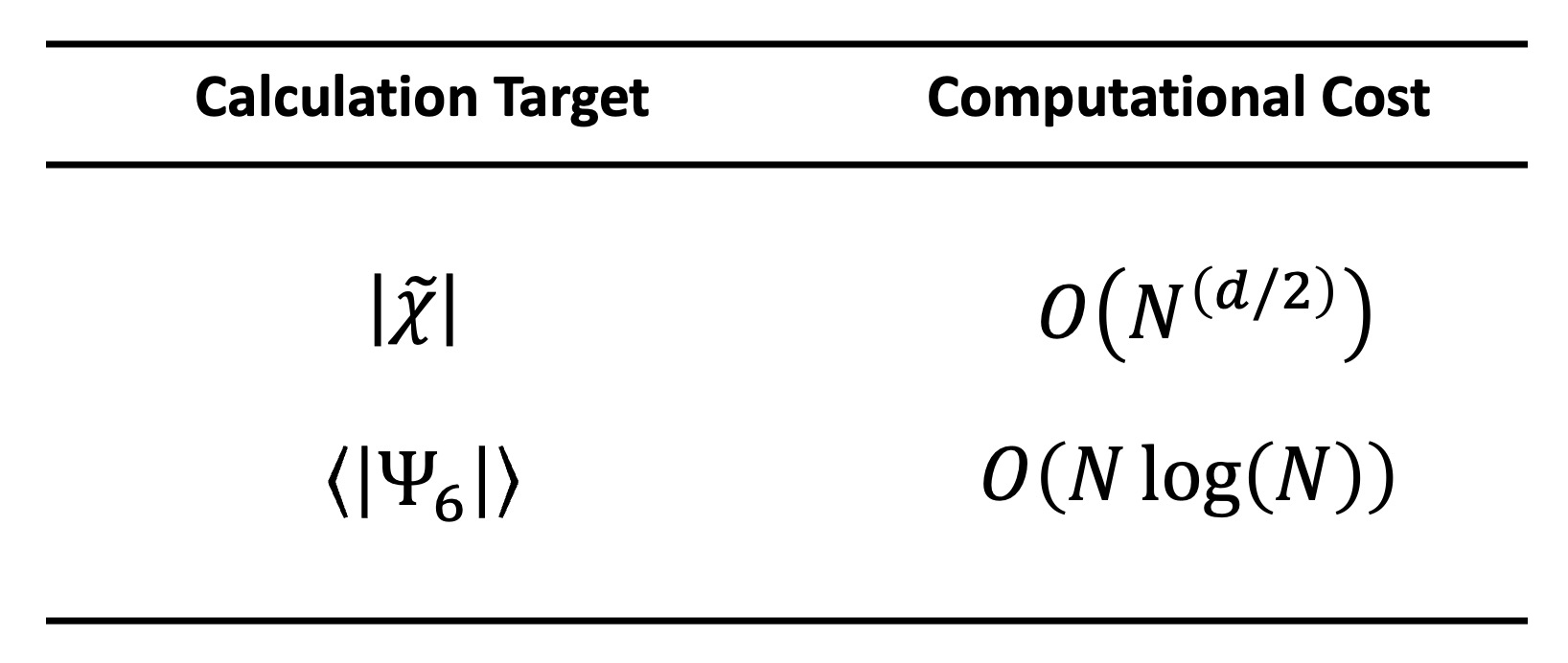}
    \label{table:data}
\end{table}

\FloatBarrier
\section*{Conclusion}
\indent
In this work, we propose a topological indicator, a Persistent Generator Count with Relative Stability (PGCRS), to 
characterize phases and phase transitions in two-dimensional quasi-particle systems.
As a model system, we use skyrmions, which can be well described as quasi-particles that form lattices in 2D systems. 
By modeling skyrmions as interacting quasi-particles and applying persistent homology (PH) to their spatial configurations, 
PGCRS selectively counts stable topological features, providing an interpretable, noise-resistant, and computationally efficient measure of lattice ordering.
It correlates with the conventional orientational order parameter $\langle|\Psi_6|\rangle$, 
and reliably traces phase transitions across solid, hexatic, and liquid states. 
Our inverse analysis reveals that persistent features in the PD with long lifetimes correspond to disordered, complex configurations, 
while short-lived features are associated with regular, ordered structures. 
The applicability of our approach is demonstrated using experimental skyrmion lattices, 
confirming that the PGCRS produces consistent and reliable phase characterization under realistic experimental conditions. 

While this work focuses on skyrmion lattice systems, the methodology is broadly applicable to other two-dimensional systems 
composed of repulsively interacting particles. 
Future extensions could incorporate insights, such as the Euler entropy, 
to further strengthen the connection between structural ordering and thermodynamic quantities \cite{ref25,ref26}.

In summary, PGCRS provides a practical and theoretically grounded framework 
for efficient, interpretable, and broadly applicable phase analysis in quasi-particle systems.

\section*{Methods}
\indent
In this work, we treat skyrmions as quasi-particles without considering their detailed spin textures,
following previous studies \cite{ref3, ref24, ref28}.
This abstraction allows us to focus on the configurational properties relevant to phase transitions, 
while significantly simplifying the computational analysis. We apply PH analysis to experimental skyrmion configurations, 
while the numerically simulated results are provided in the Supplementary information.

In the subsequent sections, we first describe the experimental setup, 
and then introduce the persistent homology analysis
and present the PH-based indicator.
Finally, we describe the conventionally used measure of the ordering. 

\subsection*{Experimental Skyrmion Lattices}
\indent
The experimental skyrmion configuration data used in this work are obtained from 
magnetic multilayer stacks: Ta(5)/Co$_{20}$Fe$_{60}$B$_{20}$(0.9)/Ta(0.07)/MgO(2)/Ta(5)~\cite{ref3,ref10}. 
Magnetic fields are applied both in-plane (IP) and out-of-plane (OOP) using an Evico Magnetics GmbH Kerr microscope, 
with the sample maintained at a constant temperature of 333.5~K. 
Skyrmions are nucleated by applying an IP field pulse under a constant OOP field. 
The lattice is subsequently equilibrated using an oscillating OOP field at 100~Hz 
(with amplitudes up to \( 60\,\upmu\mathrm{T} \)), combined with a constant OOP offset field. 
The skyrmion size is precisely controlled via the applied OOP magnetic field.
The magnetic field conditions are varied every 62.5~s (corresponding to 1000 frames) 
to gather sufficient statistics for analyzing the topological phases. 
The positions of individual skyrmions are identified using a machine-learning-based, 
pixel-wise classification algorithm~\cite{ref33}. 
The detailed analysis procedure can be found in \cite{ref10}.

\subsection*{Persistent Homology for Two-Dimensional Coordinates of Quasi-Particles}
\indent
In the persistent homology analysis, an alpha-filtration of the 0th- and
1st-degree homology is often applied to the two-dimensional coordinates of
quasi-particles \cite{edelsbrunner2002topological, zomorodian2005computing, edelsbrunner2008survey}. 
From this, the corresponding persistence diagrams (PD0 and PD1) can be generated \cite{edelsbrunner2008survey}. 
The inverse analysis method of
persistent homology is applied to the selected frames of our data,
which can identify the original structure corresponding to a specific
generator in the persistence diagram \cite{ref27}.
The details of the alpha-filtration, PDs, and inverse analysis are described in Figs.~1--3.

\subsection*{Persistent Homology-Based Indicator}
\indent
Phase transitions are accompanied by 
topology changes in the configuration manifold in some classes of systems\cite{ref15, ref25, ref26}. 
We refer the reader to Refs.\cite{ref25,ref26} for a detailed 
description of the topological invariant and its relationship with phase transitions.:

The topological invariant, Euler characteristic (EC) curve, tracks how the topology of the space changes 
as the filtration parameter \(t\) varies~\cite{ref25}, as described in the following equation:

\vspace{1em}
\begin{equation}
\chi(t) := \chi(X_{t}) = \sum_{k}^{}{( - 1)^{k}\beta_{k}(X_{t})},
\tag{6}
\end{equation}
\vspace{1em}

\noindent
where $X_{t}$ is a topological space, and \(\beta_{k}\) is \(k\)-th Betti
number, which counts the number of topological features of dimension \(k\). 
For instance, \(\beta_{0}\) and \(\beta_{1}\)correspond to the
number of connected components and loops, respectively.  
The EC is expressed as the alternating sum of Betti numbers. 
One important property of the EC is that it is a topological invariant.  
For example, if two states have different EC values, \(\chi_{1}\) and \(\chi_{2}\),  
they are topologically distinct~\cite{ref26}.  
Thus, by monitoring the evolution of the EC,  
one can detect topological transitions within the configurational space.

To summarize the EC curve and extract an integer-valued topological invariant,
we define the Persistent Generator Count with Relative
Stability, as:

\vspace{1em}
\begin{equation}
\tilde{\chi}_{\text{rel}}(X_{\text{PH}}) := \sum_{k}^{}{(-1)^{k}\beta_{k}^{\text{rel}}(X_{\text{PH}})},
\tag{7}
\end{equation}
\vspace{1em}

\noindent
where $\beta_{k}^{\text{rel}}(X_{\text{PH}})$ denotes the number of
data points in the PD that satisfy specific
stability conditions, as described in Fig.~1 and the Supplementary information. 
In this formulation, the PGCRS provides a single integer summarizing  
the stable topological features of a single persistence diagram.  
Although PGCRS is derived from the well-defined topological invariant, the EC,  
further theoretical justification is required to fully elucidate its implications.

\subsection*{Conventionally Used Measure of Ordering}

\indent
To identify the topological phases, including liquid, hexatic, and solid phases, the following quantity, 
termed the orientational correlation function, is typically used \cite{ref5, ref8, ref10, ref11, ref12},

\vspace{1em}
\begin{equation}
c_{6}(r = \left| r_{k} - r_{j} \right|) = \langle \psi_{6}(r_{k})\psi_{6}^{*}(r_{j}) \rangle, 
\tag{2}
\end{equation}
\vspace{1em}

\noindent
where
\vspace{1em}
\begin{equation}
\psi_{6}(r_{j}) = \frac{1}{n} \sum_{m=1}^{n} \exp(i6\theta_{m}^{j}), 
\tag{3}
\end{equation}
\vspace{1em}

\noindent
n is the number of nearest neighbors of the referenced particle $j$,
and $\theta_{m}^{j}$ is the angle between $r_m - r_j$ and a
fixed arbitrary axis. 
Here, the nearest neighbors are determined using
Voronoi tessellation \cite{ref30}. 
The factor 6 in the exponent reflects the sixfold
rotational symmetry of a perfect hexagonal lattice. 
When a particle's
six nearest neighbors form an ideal hexagonal arrangement, an absolute
value of $\psi_{6}$ of 1 is obtained. The KTHNY theory explains the
typical behavior of a correlation function for each topological phase,
specifically, liquid, hexatic, and solid phases exhibit exponential
decay, algebraic decay, and constant behavior close to 1 as a function
of distance $r$, respectively \cite{ref8,ref12}.

In addition to identifying the topological phase via the orientational
correlation function, another measure of the ordering is the averaged
absolute value of the local orientational order parameter $\langle|\Psi_6|\rangle$, as described in the
following equation \cite{ref5}:

\vspace{1em}
\begin{equation}
\langle|\Psi_6|\rangle = \left| \frac{1}{N} \sum_{j=1}^{N} \psi_{6}(r_{j}) \right|, 
\tag{4}
\end{equation}
\vspace{1em}

\noindent
where $N$ is the number of particles. This measure of the ordering can
be used as a quick indication of the topological phase, as the
computational cost is lower compared to the full correlation function \cite{ref5}. 
As for other reasons for using it, $\langle|\Psi_6|\rangle$ can be used as a
single scalar value, making it easier to compare across different
conditions, such as different densities, and it does not require
additional fitting, unlike the correlation function. Regarding the
fitting to the correlation function, the extracted decay value may not
be very precise, especially if noise or finite-size effects are present.

\FloatBarrier
\section*{Code Availability}
\indent
The code used in this study is available from the corresponding author upon reasonable request.

\FloatBarrier
\section*{Data Availability}
\indent
The datasets analyzed during the current study are available from the corresponding author upon reasonable request.

\printbibliography[title={References}]

\section*{Acknowledgements}
\indent
We would like to acknowledge Prof.
Hans Fangohr for his contributions during the initial phase of this work
and for his participation in regular progress discussions.

\section*{Funding}
\indent
The work was funded by the Deutsche Forschungsgemeinschaft (DFG, German
Research Foundation) projects 403502522 (SPP 2137 Skyrmionics),
49741853, and 268565370 (SFB TRR173 projects A01, B02 and A12) as well
as TopDyn, the Zeiss foundation through the Center for Emergent
Algorithmic Intelligence, the National Research Council of Science \&
Technology (NST) grant by the Korean government, MSIT (Grant No.
GTL24041-000), the Horizon 2020 Framework Program of the European
Commission under FET-Open grant agreement no. 863155 (s-Nebula) and
ERC-2019-SyG no. 856538 (3D MAGiC), and the Horizon Europe project no.
101070290 (NIMFEIA).

\FloatBarrier
\section*{Author Contributions}
\indent
M.T. performed the analysis of the simulation and experimental data. 
J.R. carried out the molecular dynamics simulations. 
R.G. conducted the Kerr microscopy measurements. 
T.B.W., C.M., M.K., and M.K. guided and supervised the research. 
All authors contributed to the interpretation of the results and the writing of the manuscript.

\FloatBarrier
\section*{Competing Interests}
\indent
The authors declare no competing interests.

\FloatBarrier
\end{document}


\section*{Supplementary Methods}

This section is organized as follows:
\begin{itemize}
  \item S1.~Criteria of Stability Condition for Persistence Diagrams
  \item S2.~Brownian Dynamics Simulation
  \item S3.~Gaussian Process Regression
\end{itemize}

\FloatBarrier
\subsection*{S1.~Criteria of Stability Condition for Persistence Diagrams}
\indent
In this work, to capture relatively stable topological features in a persistence diagram (PD), 
we apply the following selection criteria:

\vspace{1em}
\begin{align}
\beta_{0}^{\text{rel}}(X_{\text{PH}}) &:= \beta_{0}(X_t), 
\tag{8}
\\
\beta_{1}^{\text{rel}}(X_{\text{PH}}) &:= \sum_{\substack{b \leq d_{\max}^{(0)} \\ d \geq d_{\max}^{(0)} \\ (d - b) \geq \tau}} 1,
\tag{9}
\end{align}
\vspace{1em}

\noindent
where:
\begin{itemize}
  \item \(b\) and \(d\) denote the birth and death values of a generator in the PD1 diagram,
  \item \(d_{\max}^{(0)}\) is the maximum death value in the PD0 diagram,
  \item \(\tau = d_{\min}^{(0)}\) is the minimum death value in the PD0 diagram.
\end{itemize}

The minimum and maximum death values in PD0 correspond to the shortest and longest nearest-neighbor distances among skyrmions, respectively.  
The first two conditions in the selection of \(\beta_1^{\text{rel}}\) are equivalent to computing the classical Betti number \(\beta_1\) at 
a specific filtration threshold determined by the system's spatial scale.  
The third condition ensures that only persistent (i.e., topologically stable) loop-like structures are counted, where “stability” is defined 
relative to the spatial scale of a regular hexagonal configuration.

According to the law of sines, the lifetime threshold corresponding to the equilateral hexagon is given by:

\vspace{1em}
\begin{equation}
\sin\left(\frac{\pi}{n}\right) = \frac{a}{2R}
\quad \Leftrightarrow \quad
\text{lifetime} = R - \frac{a}{2} = \frac{a}{2} \left( \frac{1}{\sin\left( \frac{\pi}{n} \right)} - 1 \right),
\tag{10}
\end{equation}
\vspace{1em}

\noindent
where:
\begin{itemize}
  \item \(a\) is the side length of the regular polygon,
  \item \(R\) is the radius of the circumscribed circle (circumradius),
  \item \(n\) is the number of vertices (or sides) of the polygon,
  \item \(\text{lifetime} := d - b\) is the persistence of the topological generator.
\end{itemize}

For a regular hexagon (\(n = 6\)), the expression simplifies to \(\text{lifetime} = \frac{a}{2}\). 
Generators in PD1 with lifetimes exceeding this value are considered to reflect stable loop-like structures. 
To identify such generators, a threshold \(\tau = d_{\min}^{(0)}\) is applied, 
corresponding to the shortest nearest-neighbor distance between skyrmions.

Thus, PGCRS highlights persistent topological features consistent with the system's symmetry and 
spatial ordering.

\subsection*{S2.~Brownian Dynamics Simulation}
\indent
The skyrmion lattice states used for persistent homology analysis are
generated via Brownian dynamics simulation, specifically a Thiele
model-based approach, using the HOOMD-blue
package \cite{ref29}. 
In this work, the Magnus term is assumed negligible under high-density skyrmion conditions, 
where the resulting short mean free path limits its influence, and it is therefore omitted \cite{ref30}.
Consequently, the Thiele model simulation effectively reduces to a
standard Brownian dynamics simulation in the overdamped
limit \cite{ref30}. 
This simulation models
a two-dimensional system of quasi-particles interacting under conditions
that reproduce a range of ordering according to different phases,
including liquid, hexatic, and solid
states \cite{ref8, ref10}.

The Brownian dynamics simulation is performed using the HOOMD-blue package \cite{ref29}. 
The modelled system is a 2D system of \(N = 65000\) particles with hexagonal packing under periodic boundary conditions. 
The governing model is classical Brownian dynamics. 
Let \(\beta = 1/k_{B}T\) denote the inverse thermal energy (with $k_{B}$ the Boltzmann constant and 
$T$ the absolute temperature), $\varepsilon$ denote the interaction energy, and $\sigma$ denote the characteristic interaction range. 
We use reduced units with $k_{B}T = 1.0$ and $\varepsilon = 1.0$ (i.e., $\beta \varepsilon = 1.0$), 
and set $\sigma = 1.0$ to define the length unit. 
For a fixed potential exponent $n$, the lattice ordering depends only on 
the dimensionless density $\phi \equiv \sigma^{2}N/V$, 
where \(V\) is the area of the simulation box \cite{ref28}.
The interaction potential is given by

\vspace{1em}
\begin{equation}
U(r) = \varepsilon({\sigma}/{r})^{n}, 
\tag{1}
\end{equation}
\vspace{1em}

\noindent
where \(r\) is the interparticle distance \cite{ref3, ref5, ref28}.
In this work, we choose $n = 8.0$. 
It is worth noting that two commonly used effective skyrmion-skyrmion interaction models are 
the power-law decaying function, used in our simulations, and the exponentially decaying 
function \cite{ref3, ref5, ref10}. 
To generate states spanning a wide range of topological phases, 
22 density values are scanned from $\phi = 1.1000$ to $1.7000$, covering the entire regime from liquid to solid phases \cite{ref8}.



\subsection*{S3.~Gaussian Process Regression}
\indent
Since phase transitions are characterized by abrupt changes in system properties, 
the derivative behavior is essential for capturing variations in lattice ordering. 
To this end, we employ Gaussian Process Regression (GPR), 
which provides a statistically robust framework for generating smooth predictive curves and reliable derivative estimates, 
thereby enabling consistent comparison between PGCRS and conventional measures.
The first derivatives of the PGCRS $\langle \tilde{\chi} \rangle$ and the 
conventionally used measure of lattice ordering $\langle|\Psi_6|\rangle$ 
are estimated from a smoothed predictive curve constructed 
using GPR with a Radial Basis Function (RBF) kernel.
For the simulation data, three different RBF kernels are employed to account 
for the non-uniform sampling intervals along the density axis $\phi$. 
In contrast, the experimental dataset, which consists of more sparsely sampled 
points, is modeled using a single RBF kernel.
The details of the predictive mean curve and the associated uncertainty estimates 
from the GPR model are provided in Ref.\cite{ref34}.

\section*{Supplementary Results}
This section is organized as follows:
\begin{itemize}
  \item S4.~Gaussian Process Regression in Experimental Data
  \item S5.~Conventional Used Measure of Lattice Ordering in Simulation Data
  \item S6.~Persistence Diagrams in Simulation Data
  \item S7.~Persistence Diagram Generators and Microscopic Configurations in Simulation Data
  \item S8.~Persistent Homology-Based Indicator with Conventional Used Measure of Lattice Ordering in Simulation Data
  \item S9.~First Derivatives of Persistent Homology-Based Indicator and Conventionally Used Measure of Ordering in Simulation Data
  \item S10.~Gaussian Process Regression in Simulation Data
\end{itemize}

\subsection*{S4.~Gaussian Process Regression in Experimental Data}
\indent

\FloatBarrier
Fig. S7 presents the predicted curves as functions of the applied OOP magnetic field $B$, 
using a single Radial Basis Function (RBF) kernel applied to the experimental skyrmion lattice data. 
For $\langle \tilde{\chi} \rangle$, the length scale and variance are 21.7 and $4.53 \times 10^{-1}$, respectively. 
For $\langle|\Psi_6|\rangle$, they are 48.8 and 1.19. 
Both the predicted curves fit the data points well, further validating the consistency with each other.

\renewcommand{\thefigure}{S\arabic{figure}}
\setcounter{figure}{6}
\begin{figure}[ht]
    \centering
    \includegraphics[width=0.8\textwidth]{./Picture11.jpg}
    \caption{
        Gaussian Process Regression (GPR) predictions of $\langle \tilde{\chi} \rangle$ and $\langle|\Psi_6|\rangle$ 
        as functions of the applied OOP magnetic field $B$, using a Radial Basis Function (RBF) kernel. 
        Black dots indicate training data, blue lines show predicted means, and shaded regions represent 
        $\pm1$ standard deviation.
    }
    \label{fig:figS7}
\end{figure}

\FloatBarrier
\subsection*{S5.~Conventional Used Measure of Lattice Ordering in Simulation Data}
\indent
Fig.~S1 shows the correlation function of the local orientational order
parameter as a function of the distance for three skyrmion density values
($\phi = 1.1000$, $1.1960$, and $1.7000$) and the averaged absolute 
value of the local orientational order parameter as a function of density values. 
Each state in the left panel corresponds to a distinct phase, as indicated in Ref.\cite{ref28}. 
The correlation functions for the three density values show the exponential decay, algebraic decay,
and constant value close to 1, respectively, 
as determined by the dotted black line representing $r^{-\frac{1}{4}}$ \cite{ref10}. 
According to the KTHNY theory prediction \cite{ref8, ref10, ref12,ref28}, 
these behaviors correspond to the liquid, hexatic, and solid phases, respectively, 
with the decay rate exceeding the critical $r^{-\frac{1}{4}}$ threshold indicating the liquid phase.
The right panel shows the averaged absolute value of the local orientational
order parameter $\langle|\Psi_6|\rangle$ as a function
of density value. 
It indicates that $\langle|\Psi_6|\rangle$ increases with density, reflecting a
more ordered configurational alignment.

\setcounter{figure}{0}
\begin{figure}[ht]
    \centering
    \includegraphics[width=0.8\textwidth]{./Picture1.jpg}
    \caption{
    The correlation function of the local orientational order 
    parameter as a function of the distance for the three skyrmion density values
    $\phi = 1.1000$, $1.1960$, and $1.7000$ and 
    the averaged absolute value of the local orientational order parameter 
    as a function of density values.
    The three states correspond to the liquid, hexatic, and solid phases, respectively, 
    as determined by the dotted black line, which represents a power law decay with a critical 
    exponent $r^{-\frac{1}{4}}$ \cite{ref10}. 
    The right panel shows that $\langle|\Psi_6|\rangle$ increases with density, 
    reflecting a more ordered configurational alignment.}
    \label{fig:fig2}
\end{figure}

\FloatBarrier
\subsection*{S6.~Persistence Diagrams in Simulation Data}
\indent
Fig.~S2  displays the persistence diagrams (PDs) of the 0th- and 1st-
degree homology, respectively, generated from the three skyrmion density values
$\phi = 1.1000$, $1.1960$, and $1.7000$, corresponding to the liquid, hexatic, and solid phases. 
The "Birth" and
"Death" represent the specific time in a filtration process of the
Persistent Homology (PH) in which the topological features emerge and
disappear, respectively, and the color map represents the multiplicity
of the generators (scatter plots) in the PD. The magnitude of the values
of "Birth" and "Death" corresponds to the length of the disk located
at the coordinates of the quasi-particles, which increases during a filtration
process of the PH. 
It is worth noting that, in this work, the coordinates of the system have been normalized to eliminate a 
possible effect from the absolute density values. 
It is observed
that the distribution of generators in the PD of 0th-degree homology
(PD0) for the disordered structure (e.g., the states at skyrmion density values
$\phi = 1.1000$, and $1.1960$) is broader compared with the ordered
structure (e.g., the state at density value $\phi = 1.7000$). It is
also confirmed that the disordered structures appear to have more
generators in PD of 1st-degree homology (PD1) with a large lifetime,
which is a distance from the diagonal line in PD. 
In contrast, the ordered structure likely has fewer generators with a long lifetime in PD1.

\begin{figure}[ht]
    \centering
    \includegraphics[width=0.9\textwidth]{./Picture3.jpg}
    \caption{Average persistence diagrams (PDs) of the 0th-
    and 1st-degree homology, respectively, generated from the three skyrmion density values
    $\phi = 1.1000$, $1.1960$, and $1.7000$. The
    "Birth" and "Death" represent the specific time in a filtration
    process of persistent homology in which topological features emerge and
    disappear, respectively. The color map represents the multiplicity of
    generators (scatter plots) in the PD. Birth values of all generators in
    the 0th-degree homology (PD0) are 0, and death values in PD0 represent
    the distance between specific particle pairs, defined as distance =
    2$\sqrt{\text{death value}}$. This relationship also applies to the Birth and Death values in PD1.}
    \label{fig:fig3}
\end{figure}

\FloatBarrier
\subsection*{S7.~Persistence Diagram Generators and Microscopic Configurations in Simulation Data}
\indent
Fig.~S3 presents the inverse analysis \cite{ref27} for two skyrmion density values $\phi = 1.1000$ and $1.7000$. 
This analysis traces back the specific generators in PD to the original structure in the real-space configuration. 
In each state, the points labeled (a), (b), and (c) correspond to persistent homology generators with large lifetime
(a), small lifetime and small birth value (b), and small lifetime and large birth value (c), respectively. 
The generators with a large lifetime originate from a complex structure, and
the generators with a small lifetime originate from a simple structure.
The result confirms the previous subsection, as the
disordered structure tends to a complex alignment, while the ordered structure likely leads to a perfect alignment.
 
\begin{figure}[ht]
    \centering
    \includegraphics[width=0.8\textwidth]{./Picture4.jpg}
    \caption{Inverse analysis for the two skyrmion density values $\phi = 1.1000$ and $1.7000$, 
    tracing specific generators in PD back to real-space configurations. 
    In each state, the points labeled (a), (b), and (c) correspond to persistent homology
    generators with large lifetime (a), small lifetime and small birth value
    (b), and small lifetime and large birth value (c), respectively.}
    \label{fig:fig4}
\end{figure}

\FloatBarrier
\subsection*{S8.~Persistent Homology-Based Indicator with Conventional Used Measure 
of Lattice Ordering in Simulation Data}
\indent

Fig.~S4 presents the correlation between the Persistent Generator Count with Relative Stability
and conventionally used measure of the ordering $\langle|\Psi_6|\rangle$ and the first derivatives 
of both indicators with the Pearson correlation coefficient. 
The figure shows a positive proportional relationship, as indicated by the Pearson correlation coefficient $r = 0.981$ and $r = 0.998$.
The high sensitivity of the first derivatives supports the validity of the persistent homology-based approach, as phase 
transitions are characterized by abrupt, non-analytic changes in system properties \cite{ref39}.
The first derivatives of the Persistent Generator Count with Relative Stability and the conventionally 
used measure of the ordering are provided in Fig.~S5
The Gaussian Process Regression and the details of the first derivatives, 
used to estimate these quantities, are provided in Figs.~S6 and~S7.

\begin{figure}[ht]
    \centering
    \includegraphics[width=0.9\textwidth]{./Picture5.jpg}
    \caption{
        Relationship between the Persistent Generator Count with Relative Stability 
        $\langle \tilde{\chi} \rangle$ and 
        the conventionally used measure of the ordering $\langle|\Psi_6|\rangle$.
        The relationship between the first derivatives of both indicators is also shown 
        using Gaussian Process Regression.
        The Pearson correlation coefficient $r = 0.981$ and $r = 0.998$ indicate a good correlation.}
    \label{fig:fig5}
\end{figure}

\FloatBarrier
\subsection*{S.9~First Derivatives of Persistent Homology-Based Indicator and 
Conventionally Used Measure of Ordering in Simulation Data}
\indent

\FloatBarrier
Fig. S5 presents the Persistent Generator Count with Relative Stability 
($\langle \widetilde{\chi} \rangle$) and the conventionally used measure of the ordering 
($\langle|\Psi_6|\rangle$) as a function of the skyrmion density $\phi$. 
Both derivatives exhibit closely matching behavior, in particular around regions associated 
with phase transitions, labeled (a), (b), (c), and (d) in the inset.
As previously discussed, the consistency of the first derivatives of both indicators around phase transition regions 
supports the validity of the persistent homology-based approach, as phase transitions are characterized by abrupt, 
non-analytic changes in system properties \cite{ref39}.

\begin{figure}[ht]
    \centering
    \includegraphics[width=0.7\textwidth]{./Picture7.jpg}
    \caption{
        First derivatives of the Persistent Generator Count with Relative Stability 
        ($\langle \widetilde{\chi} \rangle$) and 
        the conventionally used measure of the ordering $\langle|\Psi_6|\rangle$ 
        as functions of density $\phi$. 
        Both first derivatives show a good agreement, especially near phase transition regions (a)-(d) 
        \cite{ref28} highlighted in the inset.}
    \label{fig:figS3}
\end{figure}

\FloatBarrier
\subsection*{S10.~Gaussian Process Regression in Simulation Data}
\indent
\FloatBarrier
Fig. S6 presents the predicted curve of the Persistent Generator Count with Relative Stability 
($\langle \widetilde{\chi} \rangle$) and the conventionally used measure of the ordering 
($\langle|\Psi_6|\rangle$) as a function of the skyrmion density $\phi$ in the simulated systems.
The predicted curves obtained using Gaussian Process Regression with composite Radial Basis Function (RBF) kernels are defined as
$\sum_{i=1}^{3} \sigma_i \exp\left(-\frac{d(\phi_j, \phi_k)^2}{2l_i^2}\right)$,
where \(l_i\) and \(\sigma_i\) are the characteristic length scales and variances of the RBF components, respectively. 
The hyperparameters have been optimized using the Gaussian Process Regression of the scikit Library \cite{ref37}, detailed as follows:

For $\langle \widetilde{\chi} \rangle$, 
\begin{itemize}
    \item $l = $ \(8.52 \times 10^{-3}\), \(3.99 \times 10^{-2}\), and \(5.13 \times 10^{-1}\), 
    \item $\sigma = $ \(6.43 \times 10^{-4}\), \(2.79 \times 10^{-3}\), and \(6.71 \times 10^{-1}\).  
\end{itemize}

For $\langle|\Psi_6|\rangle$, 
\begin{itemize}
    \item $l = $ \(8.58 \times 10^{-3}\), \(4.51 \times 10^{-1}\), and \(5.38 \times 10^{-2}\), 
    \item $\sigma = $ \(1.91 \times 10^{-3}\), \(5.90 \times 10^{-1}\), and \(1.21 \times 10^{-2}\).
\end{itemize}

\begin{figure}[ht]
    \centering
    \includegraphics[width=0.8\textwidth]{./Picture6.jpg}
    \caption{
        Predicted curves of the Persistent Generator Count with 
        Relative Stability ($\langle \widetilde{\chi} \rangle$) and the orientational order parameter 
        ($\langle|\Psi_6|\rangle$) in the simulated systems, obtained using Gaussian Process Regression (GPR). 
        The curves are shown as functions of density modeled using composite Radial Basis Function (RBF) kernels.
        Black dots represent the original simulation data points used as training inputs. 
        The blue lines indicate the GPR-predicted mean curves, and the shaded areas 
        correspond to ±1 standard deviation. 
    }
    \label{fig:figS6}
\end{figure}

\FloatBarrier
\printbibliography[title={References}]

\FloatBarrier